\documentclass{jaa}
\usepackage{natbib}
\usepackage{graphicx}
\usepackage{verbatim}
\usepackage[colorlinks, allcolors=blue]{hyperref}
\usepackage{adjustbox}
\usepackage{multirow}
\usepackage{multicol}

\begin{document}\sloppy

\title{Study of Galactic Structure Using UVIT/AstroSat Star Counts}

\author{Ranjan Kumar\textsuperscript{1*}, Ananta C. Pradhan\textsuperscript{1*}, Devendra K. Ojha\textsuperscript{2}, Sonika Piridi\textsuperscript{1}, Tapas Baug\textsuperscript{3}, and S. K. Ghosh\textsuperscript{2}} 
\affilOne{\textsuperscript{1}Department of Physics and Astronomy, National Institute of Technology, Rourkela, Odisha, 769008, India.\\}
\affilTwo{\textsuperscript{2} Tata Institute of Fundamental Research, Homi Bhabha Road, Mumbai - 400005 , India.\\}
\affilThree{\textsuperscript{3}Kavli Institute for Astronomy and Astrophysics, Peking University, 5 Yiheyuan Road, Haidian District, Beijing 100871, China\\}


\twocolumn[{

\maketitle

\corres{acp.phy@gmail.com,ranjankmr488@gmail.com}

\msinfo{04 Nov 2020}{16 Dec 2020}

\begin{abstract}
The structure of our Galaxy has been studied from ultraviolet (UV) star counts obtained with the Ultra-Violet Imaging Telescope (UVIT) on board the {\em AstroSat} satellite, in Far-UV (FUV) and Near-UV (NUV) bands. The F154W (BaF2) and N263M (NUVB4) filters were used in the FUV and NUV bands, respectively. The point sources are separated from the extra-galactic sources of UVIT observations using infrared (IR) color cut method. The observed UVIT star counts match well with the simulations obtained from the Besan\c{c}on model of stellar population synthesis towards several Galactic directions. We also estimated the scale length and scale height of the thick disc and the scale height of the thin disc using the space density function and the exponential density law for the stars of intermediate Galactic latitudes. The scale length of the thick disc ranges from 3.11 to 5.40 kpc whereas the scale height ranges from 530$\pm$32 pc to 630$\pm$29 pc. The scale height of the thin disc comes out to be in the range of 230$\pm$20 pc to 330$\pm$11 pc.

\end{abstract}

\keywords{Stars: distances - Ultraviolet: stars - Galaxy: disc}

}]


\doinum{12.3456/s78910-011-012-3}
\artcitid{\#\#\#\#}
\volnum{000}
\year{0000}
\pgrange{1--}
\setcounter{page}{1}
\lp{1}

\section{Introduction}

A major objective of modern astrophysics is to understand when and how the galaxies are formed, and how they have evolved since. 
Our own Galaxy, the Milky Way, provides a unique opportunity to study a galaxy in exquisite detail, by measuring and analyzing the properties of large samples of individual stars. The stellar population synthesis models based on star counts methods in conjunction with large area sky survey observations have considerably helped in predicting the different structural parameters of the Galaxy such as stellar densities, scale length, and scale height \citep{Gilmore1983, Robin2003, Girardi2005, Juric2008, Ivezic2012, Chen2017}. The Galaxy models are mostly based on the photometric surveys from infrared (IR) to ultraviolet (UV) bands of the electromagnetic spectrum such as massive data sets of the {\em Sloan Digital Sky Survey (SDSS)} \citep{York2000}, the {\em Two Micron All Sky Survey (2MASS)} \citep{2mass2006}, {\em Galaxy Evolution Explorer (GALEX)} \citep{Galex2005}, {\em Global Astrometric Interferometer for Astrophysics (GAIA)} \citep{Gaia2018} including millions to billions of stars.

Star counts studies in UV have made a stride after the advent of {\em GALEX}, which provided a wide sky coverage in UV allowing for a new analysis of the UV sky \citep{Bianchi2011a, Pradhan2014}. \citet{Bianchi2011a} have used the TRILEGAL model of stellar population synthesis \citep{Girardi2005} to produce UV star counts in {\em GALEX} bands and found that their model gives the closest prediction to the observed star counts. \citet{Pradhan2014} have upgraded the Besan\c{c}on model of stellar population synthesis to include the UV bands of {\em GALEX} and UVIT to produce UV star counts towards different parts of the sky. They found that the model predicted star counts match well with the observed star counts in FUV and NUV bands of {\em GALEX}. Although they developed the model to predict UV star counts in FUV and NUV filters of UVIT incorporating respective filter responses, no comparison could be carried out with observations at that time (UVIT was launched later on). Based on the model UV colors they could separate out white dwarfs (WDs) and blue horizontal branch stars (BHBs) which are evasive at other wavelength bands due to their high temperature and low luminosities. The Besan\c{c}on model of stellar population synthesis is extensively described in \citet{Robin2003, Robin2012} which produces star counts of different evolutionary stages contained in Galactic thin disc, thick disc, halo, bar and bulge. 

The thick disc density law is generally approximated by a double exponential which is a function of both scale length and scale height. The thick disc has larger scale length and scale height than the thin disc \citep{Ojha1996, Robin1996, Chen2001, Chang2011, Chen2017} although the values of these parameters are still debatable. A few studies have also revealed that the scale length of the thin disc is larger than the thick disc which contradicts to the earlier consensus \citep{Bensby2011, Cheng2012}. Even with the availability of much improved data collections, a convergence in the values of structure parameters has not yet obtained \citep{Bland2016}. 

\section{Data Reduction and Analysis}

UVIT consists of 38 cm twin telescopes; one for FUV (1300 - 1800 \AA) and the other for NUV (2000 - 3000 \AA) and visible (3200 - 5500 \AA). The light from the latter telescope is split into  NUV and visible bands using a dichroic mirror. The FUV and NUV telescopes comprise of five filters each and are operated in photon counting mode whereas the visible filters are operated in integration mode and are mostly used for tracking purpose. The field of view of both the FUV and NUV telescopes is $28'$ and the resolution is $< 1.5''$. The details about the UVIT telescopes and their calibration are reported in \cite{Tandon2017, Tandon2020}.

\begin{table}
    \centering
    \caption{The details of the UVIT observations of various Galactic fields.}
    \label{tab:observation}
    \begin{adjustbox}{width=\columnwidth, keepaspectratio}
       \begin{tabular}{c c c c c}
        \hline
        Fields & \multicolumn{2}{c}{Galactic positions} & Filters & Exposure time \\ 
        & RA (J2000) & DEC (J2000) &   & in seconds  \\
         \hline
       GAC146-46 & $29.4583^\circ$ & $13.0000^\circ$ & NUVB4 & 4,665  \\
       GC47-42 & $326.6767^\circ$ & $-8.6110^\circ$ & BaF2 & 4,989  \\
       GC47-42 & $326.6767^\circ$ & $-8.6110^\circ$ & NUVB4 & 5,521 \\
       SGP30-90 & $12.8583^\circ$ & $-27.1283^\circ$ & NUVB4 & 4,638 \\
       GC15+60 & $222.3558^\circ$ & $14.9447^\circ$ & NUVB4 & 5,718  \\
       GAC175+60 & $160.8379^\circ$ & $41.9471^\circ$ & NUVB4 & 5,694 \\
       \hline
       \end{tabular}
    \end{adjustbox}
\end{table}

We have observed five fields towards different Galactic directions such as Galactic center (GC), Galactic anti-center (GAC), and south Galactic pole (SGP) using NUVB4 filter ($\lambda_{\mathrm{eff}}$ = 2632 \AA) and one field (GC47-42, towards GC) in F154W /BaF2 filter ($\lambda_{\mathrm{eff}}$ = 1541 \AA) of UVIT. The observation details of the observed fields are given in \autoref{tab:observation}. The data reduction of the UVIT observations was performed with a customized software package CCDLAB \citep{Postma2017}. After performing all the corrections, we aligned and co-added all the orbits to obtain the final science image to perform photometry. We applied astrometry from GAIA data release 2 \citep[Gaia DR2,][]{Gaia2018} catalog using IRAF ccmap package. We were able to achieve an overall astrometric precision of $0.1''$ for our images.

We performed point spread function (PSF) photometry on the reduced final science images using DAOPHOT package \citep{Stetson1987} in IRAF\footnote{\href{https://iraf.net/}{https://iraf.net/}}. We selected 30 - 40 isolated sources for PSF modelling of the images. We got an average FWHM of 1.2$''$ for the PSF model sources on the observed images. Once the model was developed, we did an aperture photometry on NUV and FUV images. We used ALLSTAR routine to obtain the relative magnitudes of the sources in the crowded field over aperture photometry of the sources at FWHM of the PSF model stars. We performed a curve of growth analysis on PSF modeled sources to find an aperture correction value for the relative magnitudes provided by ALLSTAR routine. The aperture correction was applied to the relative magnitudes of the detected sources. Finally, we generated a catalog of UVIT observed sources by applying various selection criteria such as magnitude error cut, sharpness and chi-fit of the profile of stars. The magnitudes of the sources were corrected for extinction using $E(B-V)$ values from \citet{Schlafly2011} and then employing the extinction law of \citet{cardeli1989}.

\begin{figure*}
    \centering
    \includegraphics[width=0.495\textwidth]{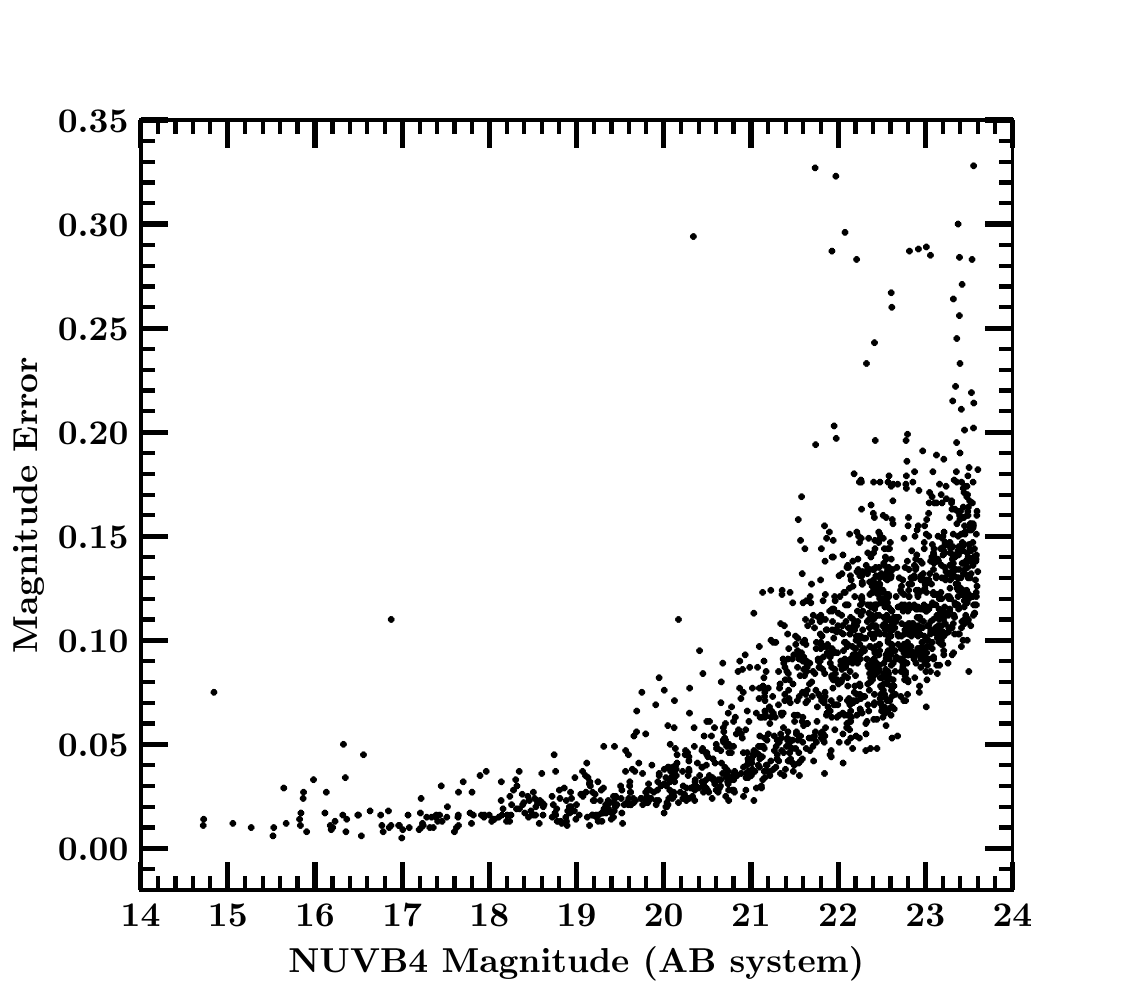}
    \includegraphics[width=0.495\textwidth]{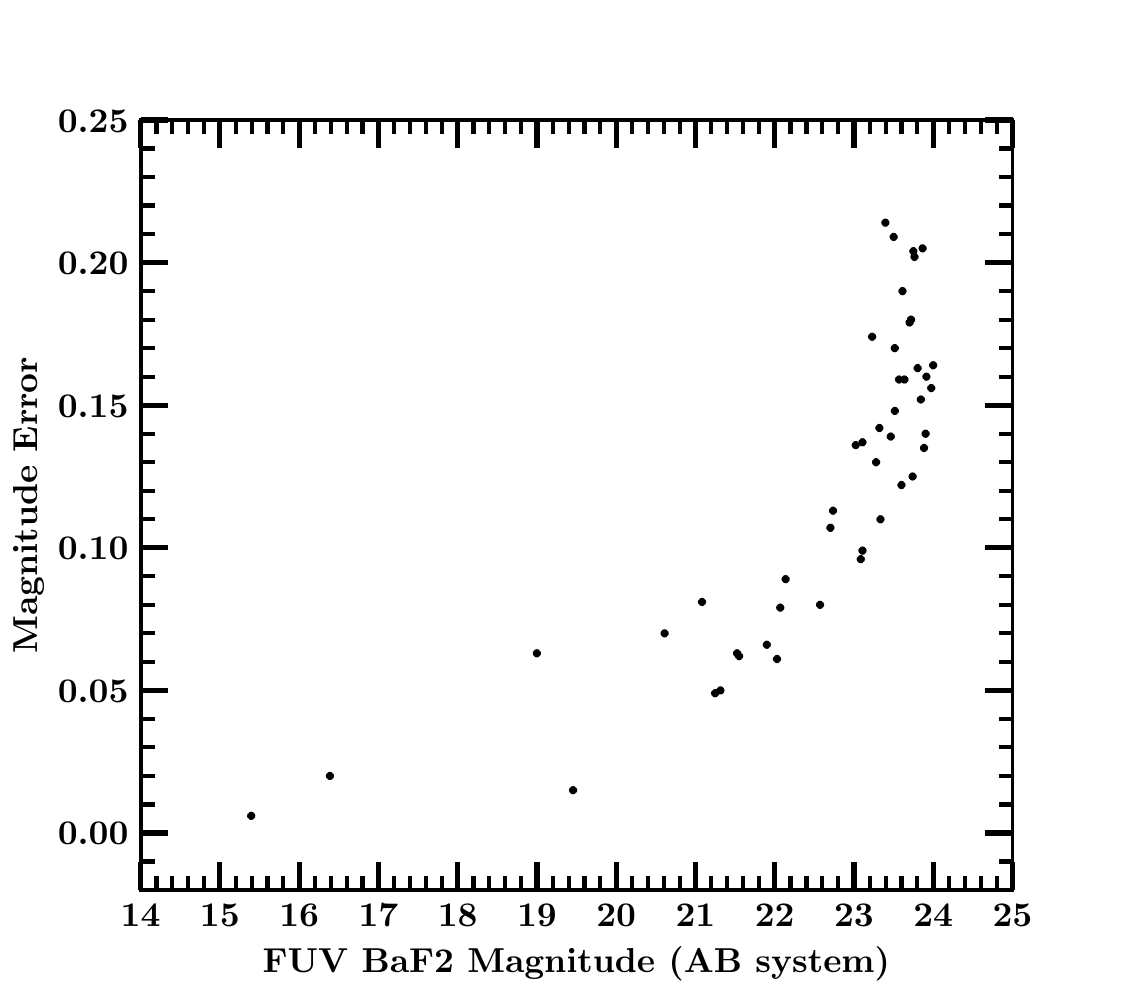}
    \caption{NUVB4 (left) and BaF2 (right) magnitudes of the observed point sources are plotted against their corresponding errors. We have retained sources with errors less than 0.2 ABmag for our analysis.}
    \label{fig:error_mag}
\end{figure*}

{\em GALEX} exposure times vary from observation to observation around the nominal exposures of 100s for all sky imaging survey (AIS). For this exposure time, the typical 5$\sigma$ detection limits for the FUV and NUV filters of {\em GALEX} AIS are $\sim$20 and $\sim$21 ABmag, respectively, and for medium imaging survey (MIS) for an exposure time of 1500s the depth is $\sim$22.7 ABmag in both FUV and NUV filters \citep{Bianchi2017}. The typical 5$\sigma$ detection limits for an exposure time of 200s for the UVIT BaF2 (FUV) and NUVB4 (NUV) wavebands are 20.0 and 21.2 ABmag, respectively. So, the typical depths reached in both {\em GALEX} and UVIT are almost similar. The magnitude error plots for UVIT FUV and NUV observed sources are shown in \autoref{fig:error_mag}. We have deep observations with exposure times more than 4.5 ks and we see that the sources upto 23.5 ABmag have the magnitude errors of less than 0.2 ABmag in both the filters. We have retained sources with errors less than 0.2 ABmag in both the NUVB4 and BaF2 filters.

\begin{figure*}
    \centering
    \includegraphics[width=0.495\textwidth]{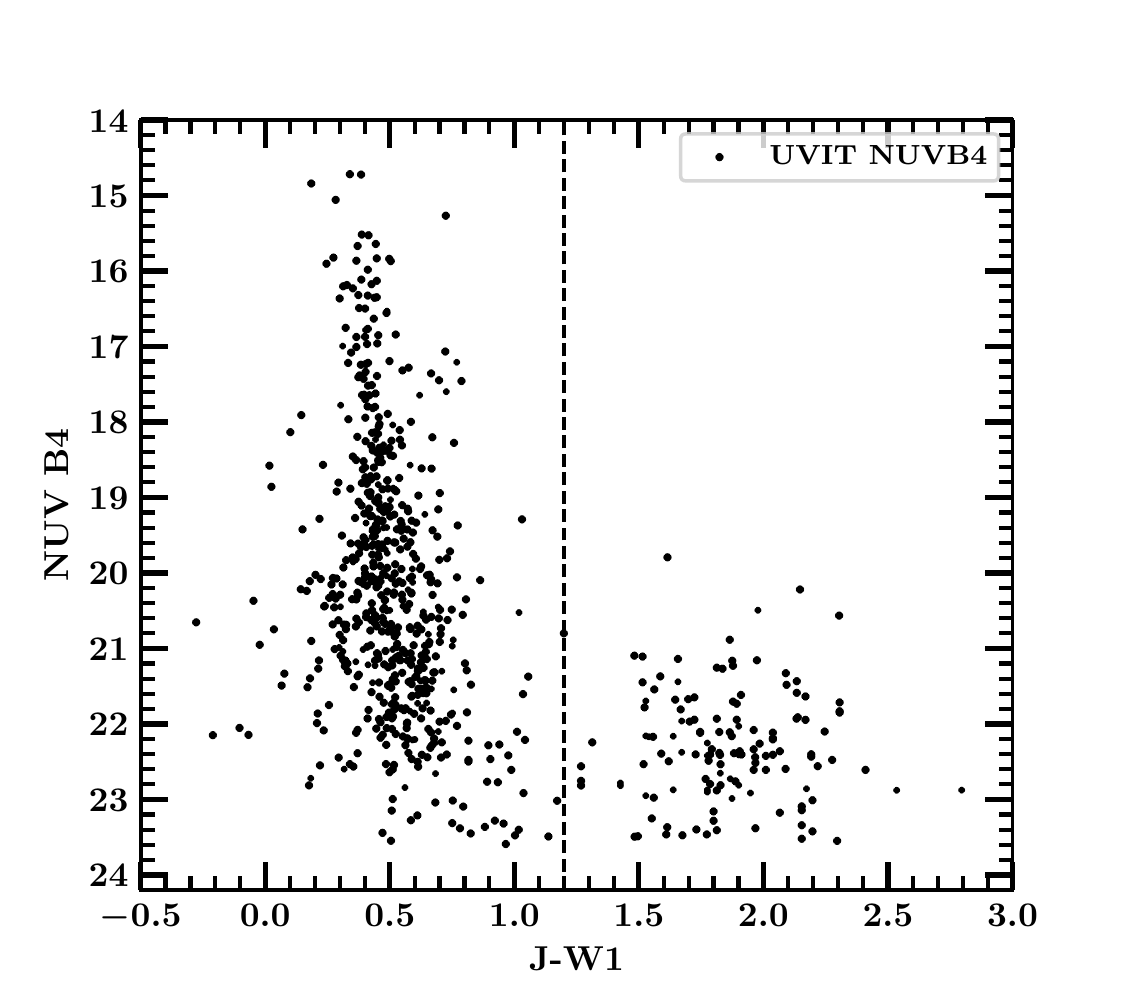}
    \includegraphics[width=0.495\textwidth]{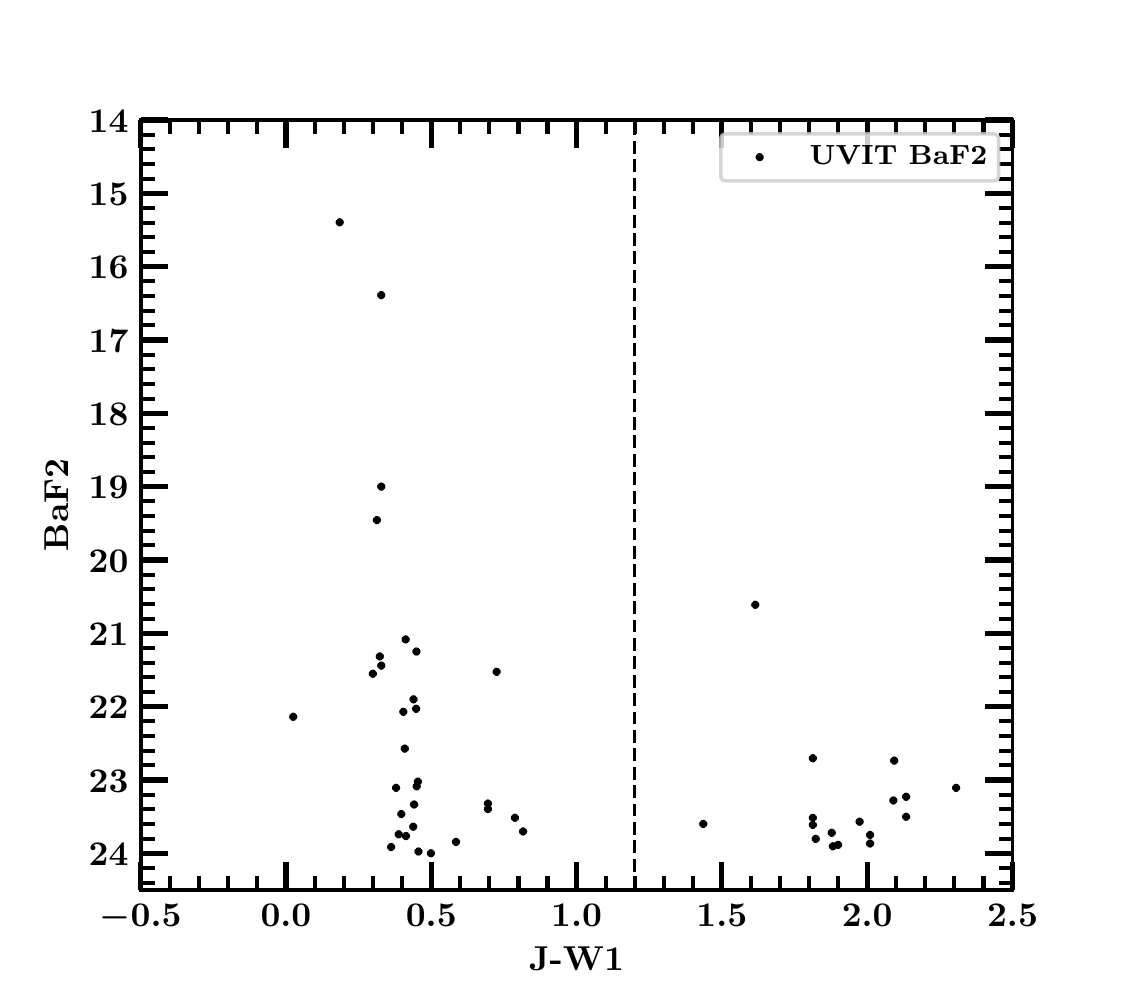}
    \caption{Color-magnitude diagrams, J$-$W1 vs NUVB4 (left) and J$-$W1 vs BaF2 (right). A vertical dashed line is drawn at the color J-W1$=$1.2 mag in the both panels. Stellar sources are well separated from the extra-galactic sources with $J-W1 < 1.2$ mag. }
    \label{fig:IR_cut}
\end{figure*}

The {\em Two Micron All Sky Survey (2MASS)} and {\em Wide-Field Infrared Survey Explorer (WISE)} observations cover almost the entire sky and our UVIT observation overlaps with {\em WISE $+$ 2MASS} survey. We cross-matched observed UVIT sources with {\em WISE $+$ 2MASS} catalog within a search radius of $3''$ using CDS X-match service available in TOPCAT software package\footnote{\href{http://www.star.bris.ac.uk/~mbt/topcat/}{http://www.star.bris.ac.uk/~mbt/topcat/}} \citep{Taylor2005}. We applied the IR color cut ($J-W1 > 1.2$ mag, where $J$ is {\em 2MASS} band at 1.24 $\mu$m and $W1$ is a {\em WISE} band at 3.4 $\mu$m) method to exclude the extra-galactic sources from our catalog \citep[see][]{Pradhan2014}. Separation of the point sources from the extra-galactic sources is clearly visible in the color-magnitude diagram as shown in \autoref{fig:IR_cut}.

\section{Comparison of model star counts}

Besan\c{c}on model is a population synthesis model developed using Galactic evolutionary scenarios and dynamics of different components of the Milky Way such as discs, bulge, halo, etc \citep{Robin2003, Robin2012}. It uses a set of evolutionary tracks, a star formation rate and an initial mass function to produce stars of different populations. \citet{Pradhan2014} have extended this model to UV passbands by including the {\em GALEX} and UVIT filters. The model has already been validated with UV star counts of {\em GALEX} catalog and also improved to generate the star counts in UVIT filters.

\begin{figure*}
    \centering
    \includegraphics[width=0.495\textwidth]{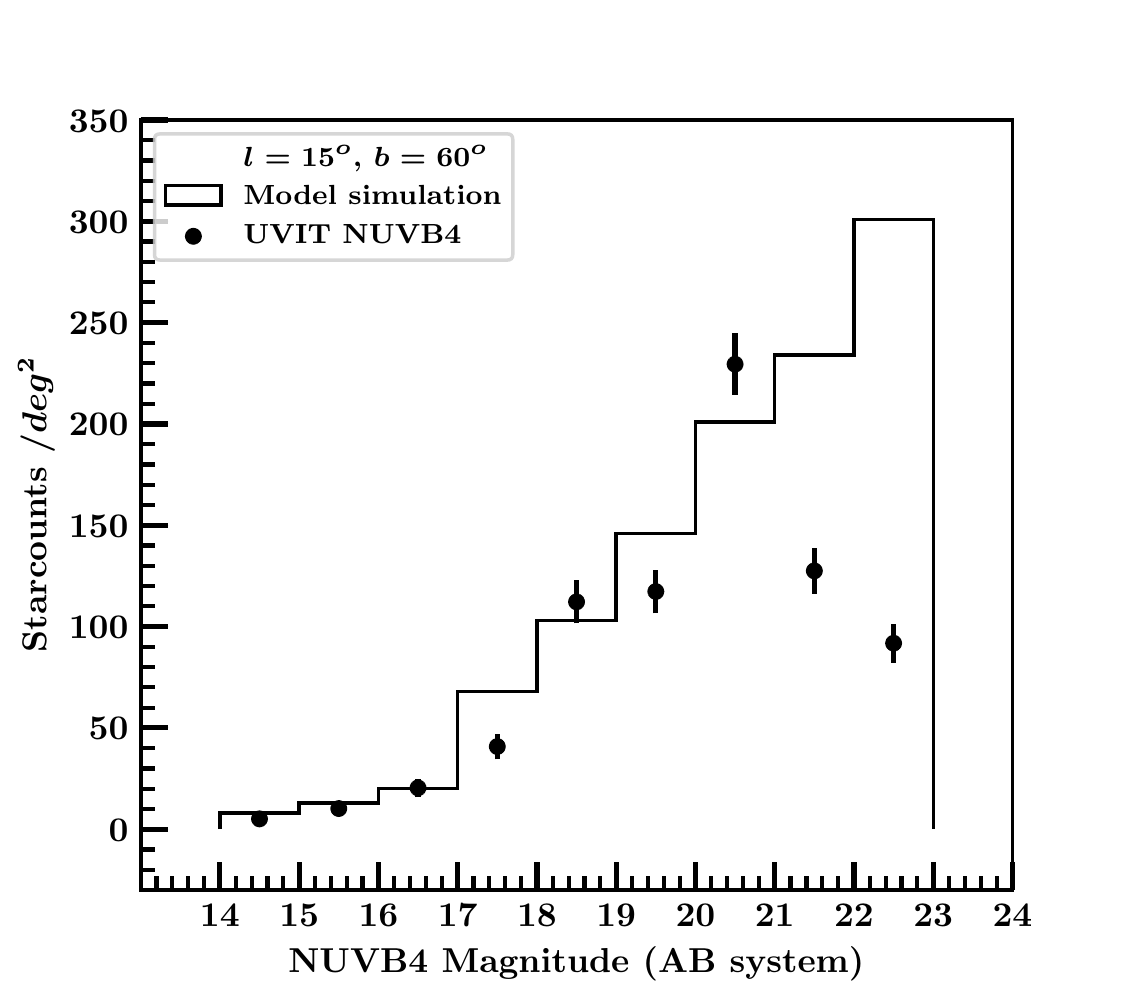}
    \includegraphics[width=0.495\textwidth]{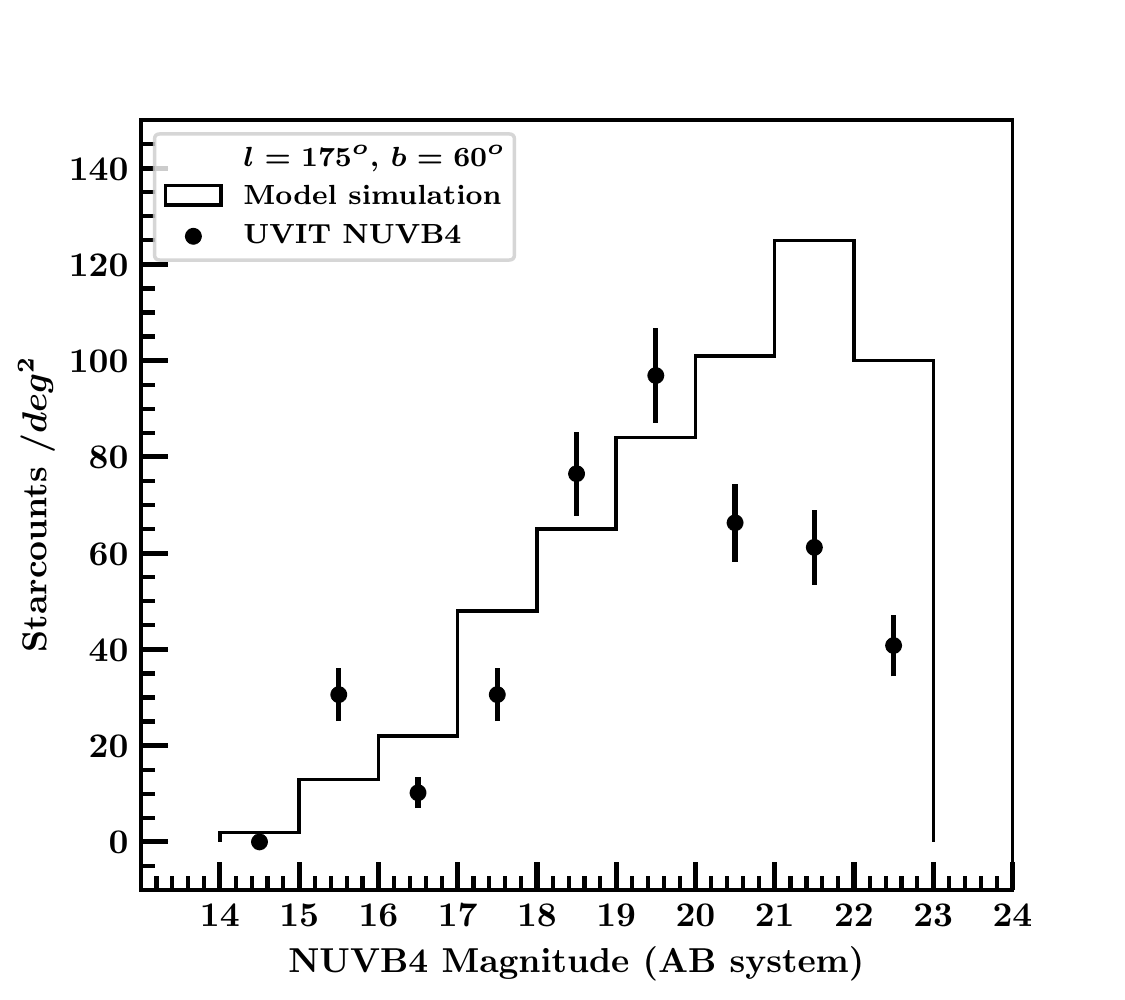}\\ 
    \includegraphics[width=0.495\textwidth]{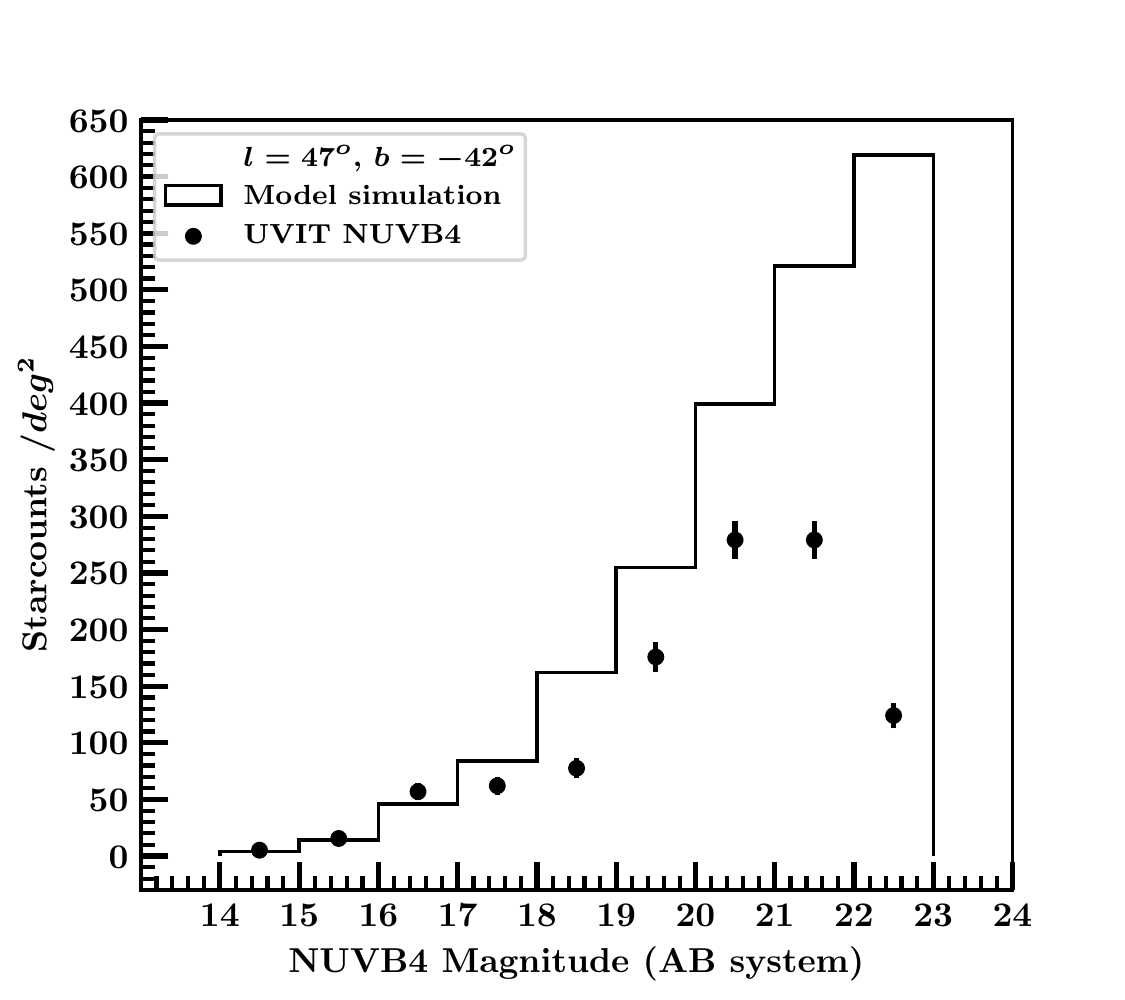} 
    \includegraphics[width=0.495\textwidth]{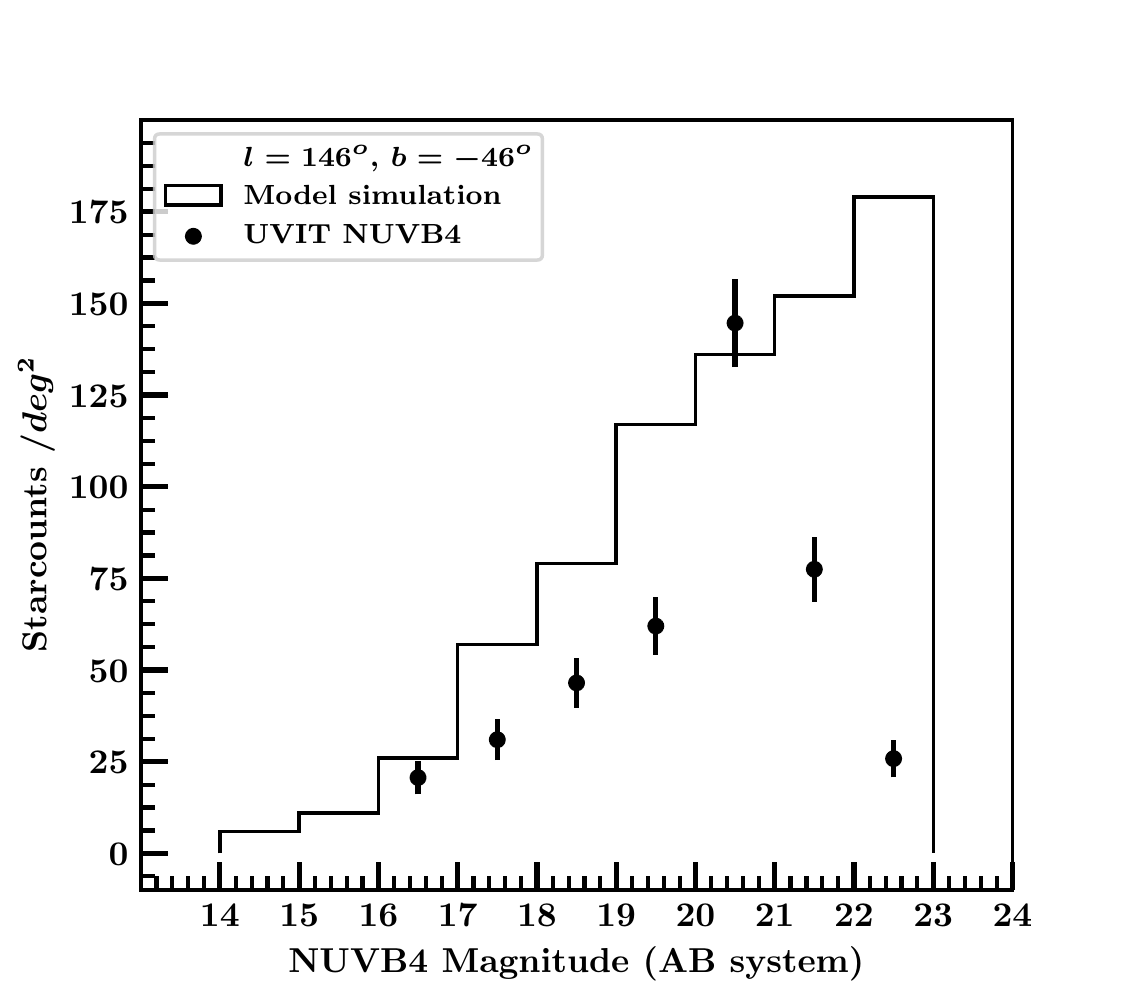} \\
    \includegraphics[width=0.495\textwidth]{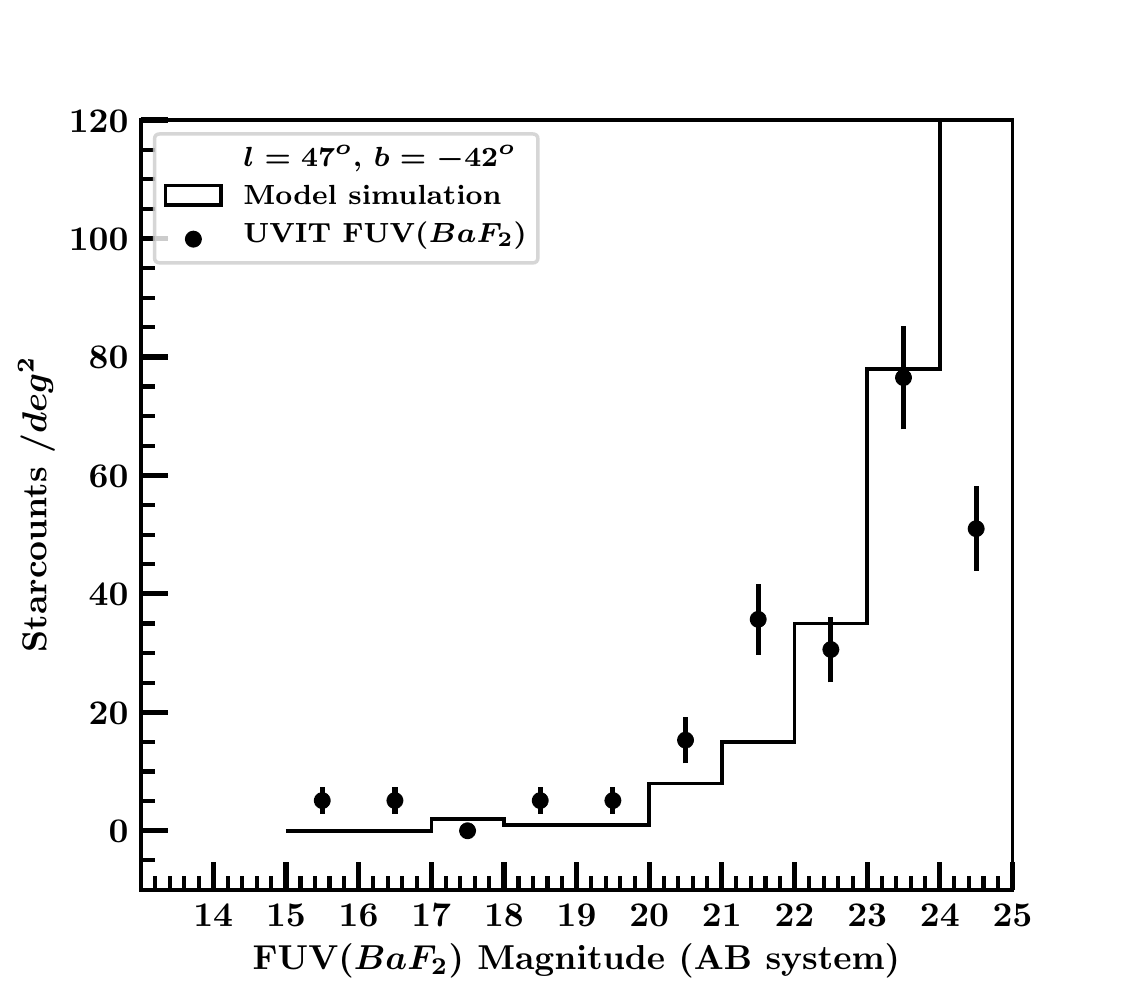} 
    \includegraphics[width=0.495\textwidth]{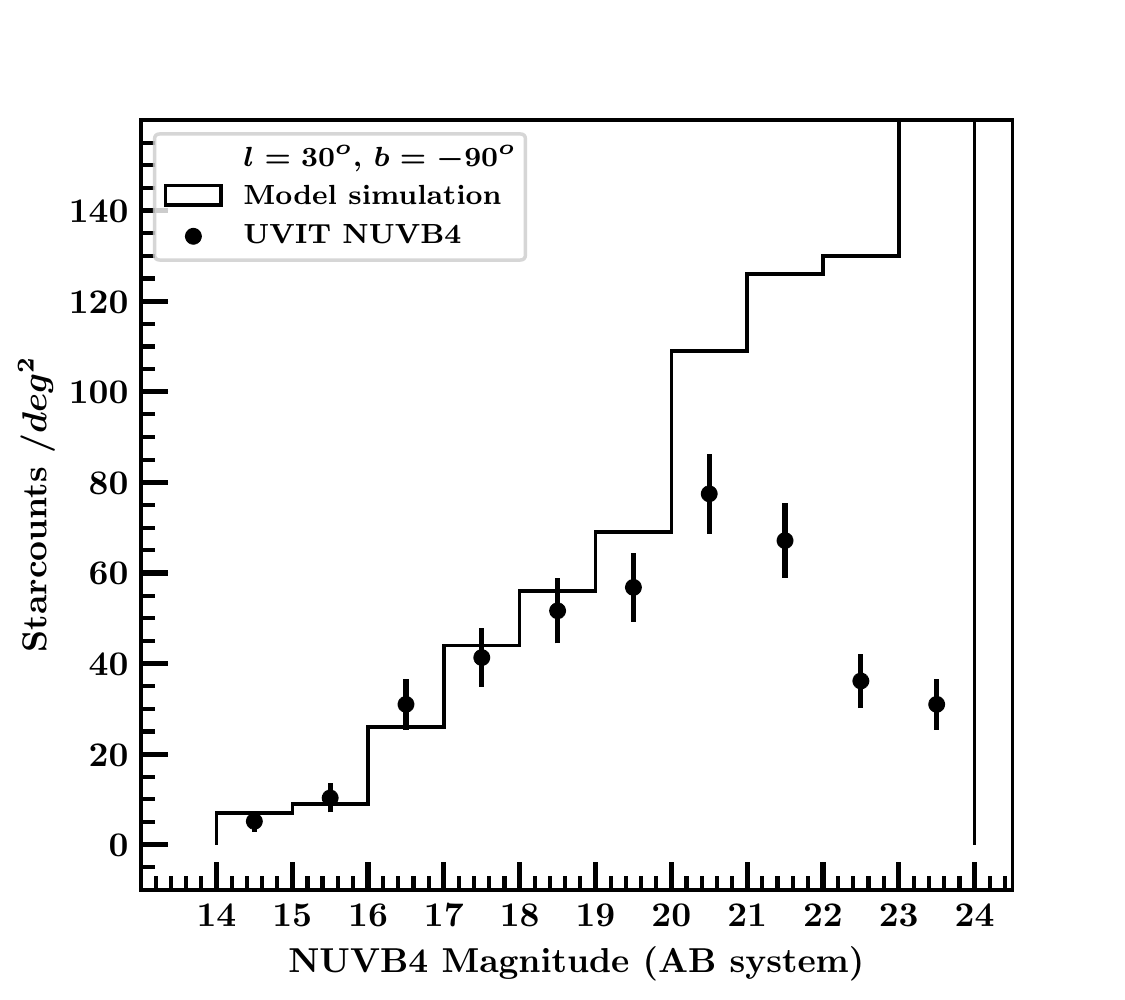} 
    \caption{Model predicted star counts are compared with the observed star counts for the NUVB4 and BaF2 filters of UVIT towards various Galactic directions. The solid circles represent the observed UV star counts along with the error bars due to Poisson noise. The solid lines represent the model generated star counts. The star counts are binned in a magnitude interval of 1.0 ABmag.}
    \label{fig:model}
\end{figure*}

Here, we verify the model star counts with our UVIT observations. We generated model simulations for the BaF2 and NUVB4 filters of UVIT in different Galactic directions. The Galactic directions were chosen in such a manner that the observation should cover regions towards GC and GAC. We have also considered the south Galactic pole direction to check the star count variation near Galactic poles. We retained all the point sources with errors less than 0.2 ABmag in both the filters. In \autoref{fig:model}, we have compared the model simulated star counts (solid line) with observations (solid circles) in UVIT NUVB4 and BaF2 filters at different Galactic latitudes. The observed and model simulated star counts were binned in a magnitude interval of 1.0 ABmag. We see that the observations in both the filters are matching with the model simulations up to their completeness limits. 

\section{Scale length and scale height of the thick disc}

Since the first evidence for the existence of the thick disc provided by \citet{Gilmore1983} from the star counts analysis, it has been established as a chemically and dynamically distinct component of the Galaxy which is substantiated by the analysis of large samples of data \citep[][and references therein]{Yong05, Bensby2011, Jacobson11, Beraldo2020}. In order to derive the structural parameters of the Galaxy (i.e, the scale length and scale height of the thin and thick discs) using star counts method, we use the density law which is approximated by a double exponential:
\begin{equation}
\label{eq:densitylaw}
\rho(R,z) = \rho(R_{0})\, exp\left(-\frac{R-R_{0}}{h_{R}}\right)exp\left(-\frac{|z|}{h_{z}}\right)
\end{equation}

Where $\rho(R_{0})$ is the normalized stellar density at the solar neighborhood, R$_{0}$ = 8.33$\pm$0.35 kpc \citep{Gillessen09} is distance of the Sun from the Galactic center, $z = d\, sin(b)$ is the height above the Galactic plane where $b$ is the Galactic latitude, $R$ is the Galactocentric distance projected on the Galactic plane, and h$_{R}$ and h$_{z}$ are the scale length and scale height of the disc, respectively. We have used star counts ratio in two galactic directions (i.e., GC and GAC) to derive the disc parameters. The star counts ratio between fields at the same Galactic latitude towards GC and GAC directions is given by,
\begin{equation}
A_{GC}/A_{GAC} = exp(+2\,|R-R_{0}|/h_{R})
\end{equation} 
Where $|R-R_{0}| =  d\, cos (b)$ is the distance of stars from the Sun on the Galactic plane. Hence, the scale length of the discs obtained from the above formula is 
\begin{equation}
\label{eq:hr}
h_{R} = \frac {2d \, cos (b)} {log(A_{GC}/A_{GAC})}
\end{equation} 

While estimating the scale length of the thick disc using \autoref{eq:hr}, we have assumed that the stellar population is homogeneous in both the Galactic directions. However, this is not exactly the case when large distances are being probed: stellar population in the inner disc, for instance, are more metal rich, hence, they could produce UV-bright sources with a different efficiency than the metal poorer populations in the outer disc. This is the kind of effect that requires a more detailed investigation, and it would be beyond the scope of this paper.

We obtained distance, $d$ of the observed UVIT sources using their parallax values from Gaia DR2 catalog \citep{Gaia2018} and then calculated the scale length of the thick disc using the UVIT observed star counts in a magnitude interval of 18.0 to 20.0 AB magnitude at GC and GAC directions at similar latitude (see \autoref{tab:density}).  

We calculated the space densities of the observed stars at two Galactic latitudes, $60^\circ$ and $-42^\circ$ in northern and southern Galactic hemispheres using the following equation:
\begin{equation}
 \centering
 \label{eq:density}
    \rho(r_1,r_2)=N_{1,2}/\Delta V_{1,2}
\end{equation}

where, $r_1$, $r_2$ are limiting distances, $N_{1,2}$ is total number of stars within distances $r_1$ and $r_2$, and partial volume $\Delta V_{1,2} = (\pi / 180)^2 (\square /3) (r_2^3 - r_1^3)$ with $\square$ being the field size in square degrees.

\begin{figure*}
    \centering
    \includegraphics[width=0.495\textwidth]{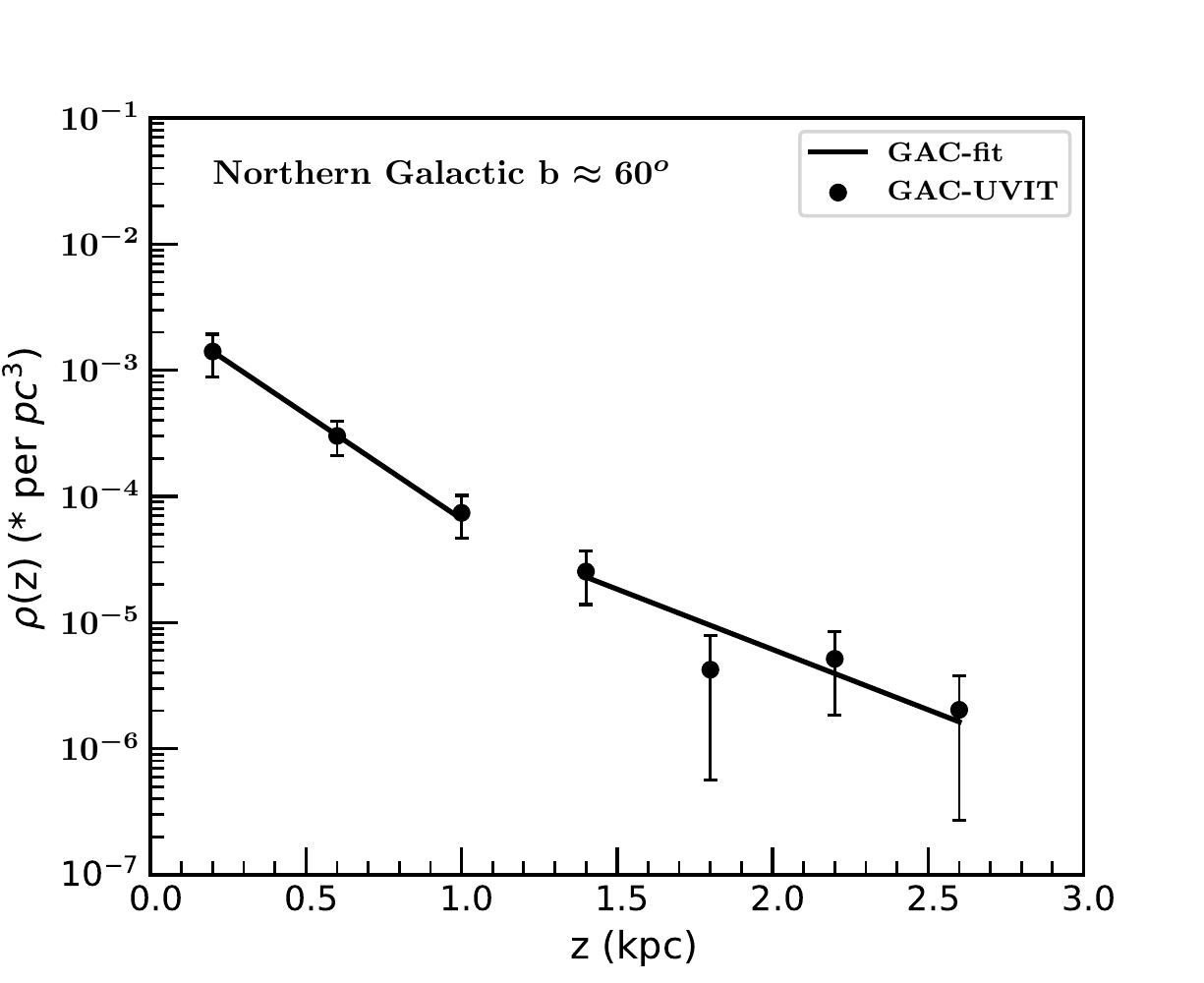}
    \includegraphics[width=0.495\textwidth]{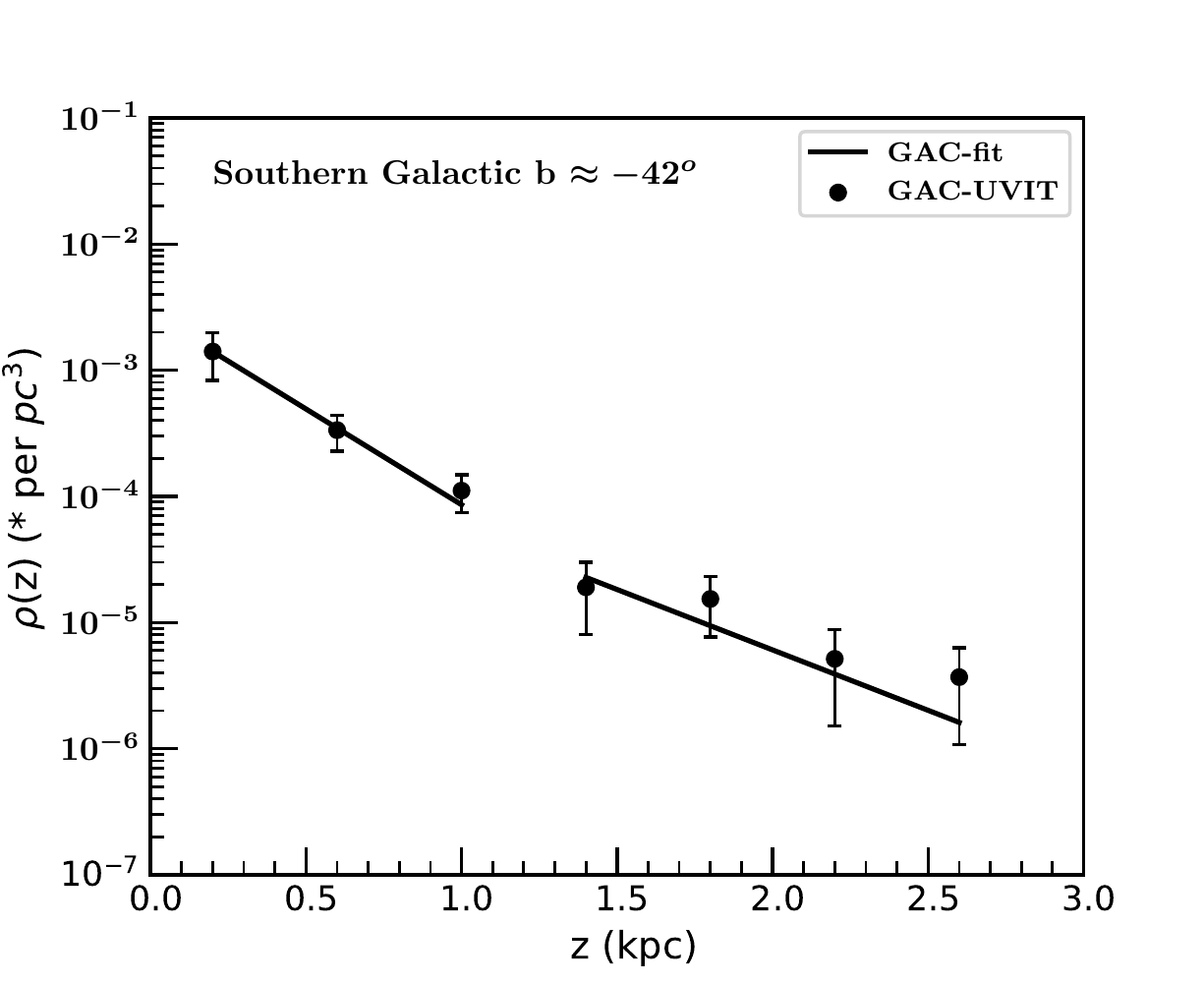}
    \caption{Space density (counts per cubic parsec) vs height above the Galactic plane (z, in kpc) for GAC fields at intermediate latitudes towards the northern (left panel) and southern (right panel) Galactic directions. Thin disc stars were fitted with an exponential density law for z $<$ 1.2 kpc, while thick disc stars were fitted from z = 1.2 to 2.8 kpc. }
    \label{fig:density}
\end{figure*}

 We then fitted the analytical density law function (solid line) to the space density (solid circles) incorporating all the associated parameters (\autoref{fig:density}). The space densities of the UVIT observed point sources towards GC and GAC directions at two latitudes $60^\circ$ and $-42^\circ$  are calculated using \autoref{eq:density}. We assumed a local density ratio for thin and thick disc stars to be 100:5 as suggested in \cite{Ojha1996} along the radial directions.  We find a turnover at z $\sim$ 1.2 kpc in the space density of stars, which shows that there are two physically distinct components: a thin disc with z $<$ 1.2 kpc and a thick disc with z $>$ 1.2 kpc. We fit the exponential density laws (\autoref{eq:densitylaw}) for thin disc stars for z $<$ 1.2 kpc and for thick disc stars from z  $\sim$1.2 to 2.8 kpc. The scale length of the thick disc and the scale heights of the thin disc and thick disc are calculated at four Galactic directions which are given in \autoref{tab:density}.

\begin{table}
    \centering
    
    \caption{Scale length and scale height of the thin and thick discs derived from UVIT star counts.}
    \label{tab:density}
    \begin{adjustbox}{width=\columnwidth, keepaspectratio}
       \begin{tabular}{c c  c c  c}
        \hline
        \multicolumn{2}{c}{Galactic positions} & \multicolumn{2}{c}{thick disc} & \multicolumn{1}{c}{thin disc}\\ \hline
        \multirow{2}{*}{$b$} & \multirow{2}{*}{$l$} &  scale length  & scale height  & scale height \\
          & & ($h_\mathrm{R}$, in kpc) & ($h_z$, in pc) & ($h_z$, in pc) \\ 
         \hline
           \multirow{2}{*}{60$^\circ$} & 15$^\circ$ & 3.11 & 570 $\pm$ 54 & 320 $\pm$ 17 \\
                                        & 175$^\circ$ & 3.11 & 650 $\pm$ 49 & 280 $\pm$ 05 \\\hline
           \multirow{2}{*}{-42$^\circ$} & 47$^\circ$ & 5.40 & 530 $\pm$ 32 & 330 $\pm$ 11 \\
                                        & 146$^\circ$ & 5.40 & 630 $\pm$ 32 & 230 $\pm$ 20 \\ \hline
        
       \end{tabular}
    \end{adjustbox}
\end{table}

A wide range of values of scale length of the thick disc and scale height of the thick disc and thin disc has been published in literature using multi-wavelength photometric surveys which demonstrates the persistence of uncertainty in these structure parameters \citep{Bland2016}. Our study of UV star counts gives a scale length of the thick disc between 3.11 kpc and 5.40 kpc, which is in close agreement with the literature values of 3.00 to 5 kpc \citep{Chen2001, Ojha01, Siegel02, Juric2008, Yaz10, Chang2011, Chen2017}. Similarly, we estimated the scale height of thick disc to be from 530 $\pm$ 32 pc to 650 $\pm$ 49 pc with several intermediate values. Our estimation matches with the estimation of the previous works; 550 - 720 pc \citep{Bilir08}, 490 - 580 pc \citep{Ak2007} and 580 - 720 \citep{Chen2001}. Our measured scale height of the thin disc ranges from 230 $\pm$ 20 pc to 330 $\pm$ 11 pc which matches with results of the recent work by \cite{Juric2008, Yaz10, Chang2011, Polido2013} and \cite{Lopez2014}.

\section{Conclusion}

We present the preliminary results of the UV star counts analysis performed using the observations obtained from UVIT on board AstroSat satellite. The Besan\c{c}on model of stellar population synthesis upgraded to produce simulations for UVIT filters is validated for two of its filters. The scale height ranges of thick disc and thin disc obtained from the UV star counts analysis are  from 530$\pm$32 pc to 630$\pm$29 pc and 230$\pm$20 pc to 330$\pm$11 pc, respectively. The scale length of the thick disc varies from 3.11 to 5.40 kpc. The values of these parameters are in well agreement with the already reported literature values.
Here, we have limited our analysis using observations through one FUV and one NUV filter of UVIT mostly covering GC and GAC regions. In the future, we will present our analysis by comparing the model predictions in other filters of UVIT towards various possible Galactic directions. This will provide us an opportunity to filter out the hot sources such as white dwarfs and blue horizontal branch stars from the sample.

\section*{Acknowledgements}
We would like to thank the referee for giving useful suggestions to improve the manuscript. We would like to thank Dr. A. C. Robin for letting us use their model of stellar population synthesis and for giving her useful inputs on the Besan\c{c}on model. RK would like to acknowledge CSIR Research Fellowship (JRF) Grant No. 09/983(0034)/2019-EMR-1 for the financial support. ACP would like to acknowledge the support by Indian Space Research Organization, Department of Space, Government of India (ISRO RESPOND project No. ISRO/RES/2/409/17-18). ACP also thanks Inter University centre for Astronomy and Astrophysics (IUCAA), Pune, India for providing facilities to carry out his work. DKO and SKG acknowledge the support of the Department of Atomic Energy, Government of India, under Project Identification No. RTI 4002. TB acknowledges the support from the National Key Research and Development Program of China (2017YFA0402702, 2019YFA0405100). This publication uses the data from the \textit{AstroSat} mission of the Indian Space Research  Organisation (ISRO), archived at the Indian Space Science Data Center (ISSDC). The UVIT data used here was processed by the Payload Operations Centre at IIA. The UVIT is built in collaboration between IIA, IUCAA, TIFR, ISRO and CSA.
\vspace{-1em}


\bibliography{draft}

\end{document}